\documentclass[11pt]{article}
\usepackage{graphicx}
\usepackage{epsfig}
\usepackage{amsmath}

\newcommand{\spone}{0.9}

\newcommand{\sptwo}{1.4}
\newcommand{\spthree}{2.4}

\newcommand{\singlespace}{\edef\baselinestretch{\spone}\Large\normalsize}

\newcommand{\doublespace}{\edef\baselinestretch{\sptwo}\Large\normalsize}
\newcommand{\threespace}{\edef\baselinestretch{\spthree}\Large\normalsize}

\singlespace
\threespace
\doublespace

\everymath={\displaystyle}
\begin{document}
\doublespace

\begin{center}
{\bf {\large Dynamics of Weyl Scale Invariant non--BPS $p=3$ Brane}\\
$~$\\
Lu-Xin Liu*\\
{\it Department of Physics, Purdue University,\\
West Lafayette, USA\\
and \\
National Institute for Theoretical Physics,\\
Department of Physics and Centre for Theoretical Physics,\\
University of the Witwatersrand,\\
Wits, South Africa\\}}
\end{center}
In this paper a Weyl scale invariant $p=3$ brane scenario is introduced, with the brane embedded in a higher dimensional bulk space with $N=1, 5D$ Super--Weyl symmetry. Its action, which describes its long wave oscillation modes into the ambient superspace and breaks the target symmetry down to the lower dimensional Weyl $W(1,3)$ symmetry, is constructed by the approach of coset method.
\begin{flushleft}
{\it
*E-mail: Luxin.Liu9@gmail.com\\}
\end{flushleft}

\pagebreak

\vspace{15pt}
\begin{flushleft}
{\Large 1. INTRODUCTION}
\end{flushleft}
\vspace{15pt}

     In the brane world scenario, our visible universe may be confined to a three-dimensional volume which resides in a higher-dimensional space, and the standard model particles are confined on this hypersurface embedded in the bulk space [1].  From the string theory point of view, we have both the BPS D--brane and non--BPS D--brane.  In the first case, the BPS D--brane carries $R-R$ charges, breaking the bulk $N=2$ spacetime supersymmetries and invariant under half of the original supersymmetries [2].  For the non--BPS case [3], there is no manifest supersymmetry for the brane world volume theory and the supersymmetry is realized as a spontaneously broken one. Actually, there have been many generalized proposals for the construction of non--BPS D--branes [4]. In [5], a space filling non--BPS D--brane, which breaks the supersymmetries of the target non--centrally extended $N=2, 3D$ superspace, is introduced by means of the Green--Schwarz approach [6]. This action, which is found to be dual to a non--BPS $p=2$ brane, can be alternatively constructed from coset approach and totally breaks the supersymmetry in the embedded $N=1, 4D$ superspace.

    On the other hand, in particle physics it has been well known that the scale symmetry plays an important role [7--11], and it has been applied to a wide range of physics sectors, which include string theory, branes [12--19], as well as recent unparticle physics theory [20]. Among them, there has been much attention on the theory of Weyl scale invariant $p$-branes, and considerable effort has been dedicated to their canonical formulations as well as their geometrical background [13, 14, 17, 18].

   According to these developments, it is the purpose of the paper to provide a straight forward scenario for the Weyl scale invariant $p$--branes. Specifically, we consider a $p=3$ brane embedded in $5D$ target bulk space with enlarged supersymmmetry, i.e. super--Weyl symmetry, which includes $N=1, 5D$ super--Poincare symmetry and the Weyl scale symmetry. As for the $p=3$ brane, it has the same three spatial dimensions as our universe appears to have. Besides, since there is no observation for the superpartners for all the particles in the standard model of particle physics, there should be no manifest supersymmetry on the brane world volume for a realistic theory. In addition, it is notable that the scale invariance is manifestly broken by the masses of the particles in the standard model; nevertheless, it is conceivable that at a much higher energy scale beyond the standard model, there are nontrivial sectors in particle physics that have the property of scale invariance. Accordingly, we consider that the dynamics of the embedded $p=3$ brane is described by Weyl scale theory, which also totally breaks the supersymmetry of the target bulk space. It is then to be hoped that progress in this direction will shed some light on the origin of the Weyl scale invariant $p$--branes and open the possibility to understand their symmetries from a dynamical point of view.

     The dynamics of the brane describes its long wave oscillation modes into the target bulk space. In this regard, we use the coset approach to realize the sponetaneous breaking of the target bulk symmetries down to the subgroup symmetries on the brane world volume. Consider the $5D$ dimension super--Weyl group $G$. Its generators include
$W(1,4)$ [7] Weyl group (formed by $M^{MN}$, $P^M$ and Weyl scale
(dilatation) generator $D$ )and eight minimum supersymmetrical charges
$Q_a$ with four complex components. The component index $a=1, 2, 3, 4$.  Its algebra has the following (anti)commutation relations:
\begin{align*}
\{Q_a ,\bar Q_b \} &= 2\gamma _{ab}^M P_M,   \tag{1} \\
[Q_a ,M^{MN} ] &= \frac{1}{2}\gamma _{ab}^{MN} Q_b,\\
 [M_{MN} ,M_{OQ} ] &= i(\eta _{NO} M_{MQ}  + \eta _{MQ} M_{NO}  - \eta
_{MO} M_{NQ}  - \eta _{NQ} M_{MO} ), \\
 [M_{MN} ,P_R ] &= i(\eta _{NR} P_M  - \eta _{MR} P_N )
\end{align*}
and
$$ [D,M_{MN} ] = 0,
 [D,D] = 0,
 [D,P_M ] =  - iP_M,  \\\eqno{(2)}
$$
$$
[D,Q_a ] =  - \frac{1}{2}iQ_a,
 [D,\bar Q_a ] =  - \frac{1}{2}i\bar Q_a,  \\
$$
where $\eta _{MN} =(1,-1,-1,-1,-1),\gamma ^M  = (\gamma ^\mu  ,\gamma ^4 )
, \gamma ^4  =  - \gamma ^0 \gamma ^1 \gamma ^2 \gamma ^3,
 \bar Q_a  =
 Q_b^ +  \gamma _{ba}^0,
 \gamma ^{MN}  = i[\gamma ^M ,\gamma ^N ]/2$
and $M,N=0,1,2,3,4$. In the present context, we consider a $p=3$ brane embedded in the $N=1, 5D$ superspace. Accordingly, the three spacial dimensional brane breaks down the target space super--Weyl invariance to a lower dimensional Weyl $W(1,3)$ symmetry, whose unbroken generators
are $\{ M_{\mu \nu } ,D,P_\mu  \}$, where the index $\mu, \nu=0,1,2,3$. As a result, the broken generators, besides the spontaneously broken automorphism generators $M_{\mu 4}$, include the translational
generator $P_4 $ transverse to the brane directions and the spinorial generators $Q_a$ and $\bar Q_a $ related to the Grassmann coordinate directions in the superspace. On the other hand, one may be tempted to consider the Lorentz group as the stability group for the total symmetry breaking. However, at the end of the paper, we point out that the infeasibility of embedding such a Minkowski brane in this super-Weyl spacetime. Since there is the presence of the dilaton field localized on the brane world volume corresponding to the target scale symmetry breaking, the dilaton field contributes a potential on the brane world volume but with an unbounded VEV. This fact then excludes the brane modes that simultaneously nonlinear realize dilatation symmetry and the supersymmetry.

\vspace{15pt}
\begin{flushleft}
{\Large 2. WEYL SCALE INVARIANT BRANE DYNAMICS}
\end{flushleft}
\vspace{15pt}

     We work on the $4D$ world volume of the submanifold. The $5D$ super Weyl algebra is to be converted to $4D$ algebra through dimension reduction. In the Weyl representation, we introduce two supersymmetrical Weyl spinor charges $Q_\alpha$ and $S_\alpha$ and define
$$
Q_a  = \frac{1}{{\sqrt 2 }}\left( {\begin{array}{*{20}c}
   {Q_\alpha  }  \\
   {i\bar S^{\dot \alpha } }  \\
\end{array}} \right) \\,
 \bar Q_a  = ( - iS^\alpha,
 \bar Q_{\dot \alpha } ) \\
\eqno{(3)}
$$
where the Weyl spinor indices $\alpha ,\dot \alpha  = 1,2$. Therefore, the algebra becomes
$$
\{ Q, \bar Q \}  = 2\sigma ^\mu  P_\mu,
 \{ S, \bar S \}  = 2\sigma ^\mu  P_\mu,   \\ \eqno{(4)}
$$
$$
 \{ Q,S\}  =  - 2\varepsilon P_4=-2 \varepsilon Z, \\
$$
$$
[K^\mu  ,Q] =  - i\sigma ^\mu  \bar S,
 [K^\mu  ,S] = i\sigma ^\mu  \bar Q, \\
$$
$$
[M^{\mu \nu } ,Q] =  - \frac{1}{2}\sigma ^{\mu \nu } Q,
 [M^{\mu \nu } ,S] =  - \frac{1}{2}\sigma ^{\mu \nu } S, \\
$$
along with their Weyl scale properties
$$
[D,K_{\mu} ] = 0,
 [D,Z] = -iZ,
 [D,P_\mu  ] =  - iP_\mu,   \\
\eqno{(5)}
$$
$$
[D,Q_\alpha  ] =  - \frac{1}{2}iQ_\alpha,
[D,S_\alpha  ] =  - \frac{1}{2}iS_\alpha,   \\
$$
where $K^\mu   = 2M^{4 \mu}$. Accordingly, from the $4D$ standpoint of view, the $N=1, 5D$ SUSY algebra is a central--charged $N=2$ four dimensional extended superalgebra with one $5D$ translation generator
becoming the central charge.

     Here, the target symmetry group $G$ is restricted to the group whose
generators can be divided into two subgroups with one is the automorphism
group of another. The transformation of the group $G$ with respect to
the coset of the unbroken automorphism generator group would give us a
description of the vierbein of the embedded submanifold, which has the same dimension as the coset space with respect to the unbroken automorphism group of the unbroken whole subgroup. In our case, the coset is $G/H$, where $H$ is formed by unbroken antomorphism generators $\{ M_{\mu \nu } ,D\} $. Hence the embedded three spatial dimension brane has $W(1,3)$ symmetry structure oscillating through the coset space formed by $G/H'$, where $H'$ is spanned by the automorphism generators $\{ M_{MN} ,D\} $ and is the automorphism group of the super spacetime group formed by the set $\{ Q_\alpha  ,\bar Q_{\dot \alpha } ,S_\alpha  ,\bar S_{\dot \alpha } ,P_\mu  \} $. The submanifold which the brane sweeps out has the dimensions of the coset $G'/H$ with tangent space group $H$,
where $G'$ is spanned by the unbroken automorphism generators $\{ M_{\mu \nu } ,D\}$ and the unbroken spacetime translation generators $P_{\mu}$. The group elements of $G$ can be parameterized in some neighborhood of the identity element as follows
$$
g = e^{i[a^\mu  p_\mu   + \xi Q + \bar \xi \bar Q + \psi S + \bar \psi
\bar S + zZ + b^\mu  K_\mu   + \alpha ^{\mu \nu } M_{\mu \nu }  + dD]}.  \\
\eqno{(6)}
$$
We choose static gauge for the parameterization of the brane world volume, then the space time coordinates $x^{\mu}$ lie in the brane parameterizing direction $\xi^{\mu}$. The coset $G/H$ representative elements then have the form
$$
\Omega  = e^{ix^\mu  p_\mu  } e^{i\phi (x)Z} e^{i[\theta (x)Q + \bar
\theta (x)\bar Q + \lambda (x)S + \bar \lambda (x)\bar S]} e^{iu^\mu
(x)K_\mu  }.  \\
\eqno{(7)}
$$
in which $\phi (x),
 \theta (x), \bar \theta (x),
 \lambda (x),
 \bar \lambda (x)$ and
 $u(x)$ are the Nambu--Goldstone fields corresponding to each broken generators. Since the elements of group $G$ can be decomposed uniquely into a product form, given by a coset representative element and a subgroup element $h$, the transformations of these Nambu--Goldstone fields can be obtained by acting a group operation on $\Omega$ from the left, i.e.
$$
g\Omega  = \Omega 'h,  \\
\eqno{(8)}
$$
where the new coset element
$
\Omega ' = e^{ix'^\mu  p_\mu  } e^{i\phi '(x')Z} e^{i[\theta '(x')Q + \bar
\theta '(x')\bar Q + \lambda '(x')S + \bar \lambda '(x')\bar S]}
e^{iu'^\mu  (x')K_\mu  }.
$

     As for the spontaneous symmetry breaking of (super)spacetime, the
Nambu--Goldstone fields are those associated with the broken
(super)translation operators, and the superfluous Nambu--Goldstone field $u^\mu$ can be eliminated by imposing covariant constraints on the Cartan differential one forms. Explicitly, the Cartan one-forms can be expanded with respect
to the full generators as follows
\begin{align*}
\Omega ^{ - 1} d\Omega = &e^{ - iu^\mu  (x)K_\mu  }e^{ - i[\theta (x)Q +
\bar \theta (x)\bar Q + \lambda (x)S + \bar \lambda
(x)\bar S]}e^{ - i\phi (x)Z}e^{ - ix^\mu  p_\mu  } \cdot \tag{9} \\
& d(e^{ix^\mu  p_\mu  } e^{i\phi (x)Z} e^{i[\theta (x)Q + \bar
\theta (x)\bar Q + \lambda (x)S + \bar \lambda (x)\bar S]} e^{iu^\mu
(x)K_\mu  } ) \\
=&i(\omega ^a p_a  + \omega _Q^\alpha  Q_\alpha   + \bar \omega _{\bar
Q\dot \alpha } \bar Q^{\dot \alpha } +\omega _Z Z + \omega _S^\alpha  S_\alpha   + \bar \omega _{\bar S\dot
\alpha } \bar S^{\dot \alpha }  \\
+& \omega _k^a K_a  + \omega _M^{\mu \nu }
M_{_{\mu \nu } }  + \omega _D D).
\end{align*}
According to Eq.(8), the Cartan one-forms transform as
$$
{\Omega'} ^{ - 1} d{\Omega'}=h(\Omega ^{ - 1} d\Omega)h^{-1}+hdh^{-1}
\\
\eqno{(10)}
$$

By using
the differentiation formula for exponent $\exp ( - b)d\exp (b) =
\sum\limits_{k = 0}^\infty  {\frac{{( - 1)^k
}}{{(k + 1)!}}} (ad_b )^k db $, where $ad_b (a) = [b,a] $ is the adjoint
operation,
we get the following building blocks related to the construction of the effective brane action, which is invariant under $G$ transformations:
\begin{align*}
\omega ^a  = &(dx^b  + i\theta \sigma ^b d\bar \theta  - id\theta \sigma
^b
\bar \theta  + i\lambda \sigma ^b d\bar \lambda  - id\lambda \sigma ^b
\bar \lambda ) \tag{11}  \\
  &\cdot (\delta _b^{\hspace{3pt} a}  + (\cosh 2\sqrt {u^2 }  -
1)\frac{{u_b u^a
}}{{u^2 }}) - (d\phi  + id(\theta \lambda  - \bar \theta \bar \lambda
)\frac{{\sinh 2\sqrt {u^2 } }}{{\sqrt {u^2 } }}u^a,  \\
 \omega _Z  = &(d\phi  + id(\theta \lambda  - \bar \theta \bar \lambda
))\cosh 2\sqrt {u^2 }  \\
  &- (dx^a  + i(\theta \sigma ^a d\bar \theta  - d\theta \sigma ^a \bar
\theta  + \lambda \sigma ^a d\bar \lambda  - d\lambda \sigma ^a \bar
\lambda ))\frac{{\sinh 2\sqrt {u^2 } }}{{\sqrt {u^2 } }}u_a,  \\
 \omega _D  = &0,
\end{align*}
where $a=0,1,2,3$. We use Latin letters $a,b…$ to represent the tangent spacetime indices, and Greek letters $\mu,\nu…$ to denote $1+3$ general coordinate indices in what follows. The induced vierbein can be found by expanding the Cartan one-forms associated with the unbroken spacetime generators with respect to the general coordinate differentials $dx^{\mu}$, i.e. $\omega ^a  = dx^\mu e_\mu ^{\\\ a} $. The spin connection $\omega _{\rho M}^{\mu \nu }$ can also be constructed in a similar way, i.e., $\omega _M^{\mu \nu }=dx^{\rho}\omega _{\rho M}^{\mu \nu }$. Considering the unbroken subgroup $W(1,3)$ (Weyl group) and Eq.(10), it follows that the dilatation transformation property of the tangent and general coordinates takes the form
$$
\omega ^a  \to e^d \omega ^a \hspace{3pt} or \hspace{3pt} dx^{_\mu}  \to e^d dx^{_\mu}, \\
\eqno{(12)}
$$
then it is concluded that they have scale dimension $1$. On the brane world volume, the interval is
$$
ds^2  = g_{\mu \nu } dx^\mu  dx^\nu   = \eta _{ab} e_\mu  ^{\\\ a}e_\nu^{\\\ b} dx^\mu  dx^\nu,
\\
\eqno{(13)}
$$
in which the metric tensor is given by
$$
g_{\mu \nu }  = e_\mu^{\\\ a} e_\nu^{\\\ b} \eta _{ab}.\\
\eqno{(14)}
$$
Besides, under the scale transformation, in the local tangent space we have the transformation
$$
x^a \to x'^a  = e^{d} x^a
\\
\eqno{(15)}
$$
it follows that the interval transforms as
$$
ds^2 \to ds'^2  = e^{2d} ds^2.
\\
\eqno{(16)}
$$
This can also be concluded from Eqs.(12) and (13). Explicitly, in accord with Eq.(11), the vierbein, which is not an independent variable, has the following form
\begin{align*}
e_\mu  ^{\hspace{3pt} a}= &(\delta _\mu  ^{\hspace{3pt} b}  + i\theta
\sigma ^b \partial _\mu  \bar
\theta  - i\partial _\mu  \theta \sigma ^b \bar \theta  + i\lambda \sigma
^b \partial _\mu  \bar \lambda  - i\partial _\mu  \lambda \sigma ^b
\lambda )  \tag{17}   \\
  &\cdot (\delta _b^{\hspace{3pt} a}  + (\cosh 2\sqrt {u^2 }  -
1)\frac{{u_b u^a
}}{{u^2 }}) - \partial _\mu  (\phi  + i\theta \lambda  - i\bar \theta \bar
\lambda )\frac{{\sinh 2\sqrt {u^2 } }}{{\sqrt {u^2 } }}u^a  \\
  = &A_\mu  ^{\hspace{3pt} b}  \cdot (\delta _b^{\hspace{3pt} a}  + (\cosh
2\sqrt {u^2 }  -
1)\frac{{u_b u^a }}{{u^2 }} - \tilde D_b (\phi  + i\theta \lambda  - i\bar
\theta \bar \lambda )\frac{{\sinh 2\sqrt {u^2 } }}{{\sqrt {u^2 } }}u^a ),
\end{align*}
in which $A_\mu  ^{\\\ b}  = (\delta _\mu  ^{\\\ b}  + i\theta \sigma ^b
\partial _\mu
\bar \theta  - i\partial _\mu  \theta \sigma ^b \bar \theta  + i\lambda
\sigma ^b \partial _\mu  \bar \lambda  - i\partial _\mu  \lambda \sigma ^b
\lambda )$, and $\tilde D_b  = A_b^{ - 1\mu } \partial _\mu$
is the Akulov--Volkov derivative[21--24]. Imposing the
covariant condition $\omega_Z=0$ on the Cartan one-forms, as a result
of the inverse Higgs Mechanism [25], the field $u_m$ can be eliminated by the following relation
$$
u_b \frac{{\tanh 2\sqrt {u^2 } }}{{\sqrt {u^2 } }} = \tilde D_b \phi  +
i\tilde D_b \theta \lambda  + i\theta \tilde D_b \lambda  - i\tilde D_b
\bar \theta \bar \lambda  - i\bar \theta \tilde D_b \bar \lambda  = \tilde
D_b \Phi
\eqno{(18)}
$$
where $\Phi  = \phi  + i\theta \lambda  - i\bar \theta \bar \lambda$. It can further yield
$$
\cosh 2\sqrt {u^2 }  = \sqrt {\frac{1}{{1 - \tanh ^2 2\sqrt {u^2 } }}}  =
\sqrt {\frac{1}{{1 - (\tilde D\Phi )^2 }}},  \\
\eqno{(19)}
$$
with
$$
\tanh ^2 2\sqrt {u^2 }  = (\tilde D\Phi )^2.
\eqno{(20)}
$$
Plugging Eq.(18) back into Eq.(17), the vierbein hence has the simple form
\begin{align*}
e_\mu  ^{\hspace{3pt} a} = &A_\mu  ^{\hspace{3pt} b}  \cdot (\delta
_b^{\hspace{3pt} a}  + \frac{{1 - \cosh 2\sqrt
{u^2 } }}{{\cosh 2\sqrt {u^2 } }}\frac{{u_b u^a }}{{u^2 }})   \tag{21}   \\
  = & A_\mu  ^{\hspace{3pt} b}  \cdot (\delta _b^{\hspace{3pt} a}  +
\frac{{1 - \cosh 2\sqrt {u^2 }
}}{{\cosh 2\sqrt {u^2 } }}\frac{{\tilde D_b \Phi \tilde D^a \Phi }}{{\tanh
^2 2\sqrt {u^2 } }}) \\
  = & A_\mu  ^{\hspace{3pt} b}  \cdot (\delta _b^{\hspace{3pt} a}  +
\frac{{\tilde D_b \Phi \tilde
D^a
\Phi }}{{(\tilde D\Phi )^2 }}(\sqrt {1 - (\tilde D\Phi )^2 }  - 1)).
\end{align*}
The metric tensor hence becomes
$$
g_{\mu \nu }  = e_\mu ^{\\\ a} e_\nu ^{\\\ b} \eta _{ab}  = A_\mu  ^{\\\
a} A_\nu
^{\\\ b} \eta _{ab}  - \partial _\mu  \Phi \partial _\nu  \Phi.  \\
\eqno{(22)}
$$
Introduce five dynamic variables $ X^M  = (X^a ,X^4 ) = (X^a ,\Phi ) $,
which are defined as follows
\begin{align*}
dX^a  = &dx^\mu  A_\mu ^{\hspace{3pt}a}; \tag{23} \\
dX^4  = &\partial _\mu  (\phi  + i\theta \lambda  - i\bar \theta \bar
\lambda )dx^\mu.
\end{align*}
Therefore, in the static gauge $\xi ^\mu   = x^\mu$, the metric tensor in
Eq.(22) gives us
\begin{align*}
g_{\mu \nu }  = &\eta _{MN} \frac{{\partial X^M }}{{\partial \xi ^\mu
}}\frac{{\partial X^N }}{{\partial \xi ^\nu  }}  \tag{24} \\
  = &\eta _{\mu \nu }  + i\theta \sigma _\nu
\mathord{\buildrel{\lower3pt\hbox{$\scriptscriptstyle\leftrightarrow$}}
\over \partial } _\mu  \bar \theta  + i\lambda \sigma _\nu
\mathord{\buildrel{\lower3pt\hbox{$\scriptscriptstyle\leftrightarrow$}}
\over \partial } _\mu  \bar \lambda  + i\theta \sigma _\mu
\mathord{\buildrel{\lower3pt\hbox{$\scriptscriptstyle\leftrightarrow$}}
\over \partial } _\nu  \bar \theta  + i\lambda \sigma _\mu
\mathord{\buildrel{\lower3pt\hbox{$\scriptscriptstyle\leftrightarrow$}}
\over \partial } _\nu  \bar \lambda  \\
  &+ (i\theta \sigma _b
\mathord{\buildrel{\lower3pt\hbox{$\scriptscriptstyle\leftrightarrow$}}
\over \partial } _\mu  \bar \theta  + i\lambda \sigma _b
\mathord{\buildrel{\lower3pt\hbox{$\scriptscriptstyle\leftrightarrow$}}
\over \partial } _\mu  \bar \lambda ) \cdot (i\theta \sigma ^b
\mathord{\buildrel{\lower3pt\hbox{$\scriptscriptstyle\leftrightarrow$}}
\over \partial } _\nu  \bar \theta  + i\lambda \sigma ^b
\mathord{\buildrel{\lower3pt\hbox{$\scriptscriptstyle\leftrightarrow$}}
\over \partial } _\nu  \bar \lambda ) \\
  &- (\partial _\mu  \phi  + i\partial _\mu  \theta \lambda  + i\theta
\partial _\mu  \lambda  - i\partial _\mu  \bar \theta \bar \lambda  -
i\bar \theta \partial _\mu  \bar \lambda ) \\
  &\cdot (\partial _\nu  \phi  + i\partial _\nu  \theta \lambda  + i\theta
\partial _\nu  \lambda  - i\partial _\nu  \bar \theta \bar \lambda  -
i\bar \theta \partial _\nu  \bar \lambda ) \\
  = & \eta _{MN} \frac{{\partial x^M }}{{\partial \xi ^\mu
}}\frac{{\partial x^N }}{{\partial \xi ^\nu  }} +
\theta,
\bar
\theta, \lambda, and {\hspace{3pt}} \bar \lambda {\hspace{3pt}} terms,
\end{align*}
where $x^M  = (x^0 ,x^1 ,x^2 ,x^3 ,\phi )$. Consequently, in contrast
with the normal induced metric $g_{\mu \nu }  = \eta _{MN} \frac{{\partial x^M }}{{\partial \xi ^\mu  }}\frac{{\partial x^N }}{{\partial \xi ^\nu  }}$ on the $p$--brane world volume, there are modification terms to the metric, which are contributed by the Nambu--Goldstone fields $\theta(x), \bar \theta(x), \lambda(x)$, and $\bar \lambda(x)$, as a result of the broken symmetries associated with the superspace coordinate directions.

   Introduce an auxiliary (intrinsic) metric $\rho _{\mu \nu }$ on the $p=3$ brane world volume, whose scale dimension is 2 as induced by the scale transformation $\xi ^\mu \to e^d \xi^\mu$, i.e.
$$\rho _{\mu \nu
}  \to e^{2d} \rho _{\mu \nu }
\\
\eqno{(25)}
$$
or
$$d^4x \sqrt {\left| \rho  \right|}
 \to d^4x'\sqrt {\left| {\rho '} \right|}  = e^{4d} d^4x \sqrt {\left|
\rho  \right|}.
\\
\eqno{(26)}
$$
By using Eq.(24), the effective scale invariant action of the $p=3$ brane world volume is secured as follows
\begin{align*}
I =&  - T\int {d^4 x} \sqrt {\left| \rho  \right|} [\frac{1}{4}\rho ^{\mu
\nu } \eta _{MN} \partial _\mu  X^M \partial _\nu  X^N ]^2  \tag{27} \\
  =&  - T\int {d^4 x} \sqrt { - \rho } [ \frac{1}{4}\rho ^{\mu \nu }
(\eta
_{\mu \nu }  + i\theta \sigma _\nu
\mathord{\buildrel{\lower3pt\hbox{$\scriptscriptstyle\leftrightarrow$}}
\over \partial } _\mu  \bar \theta  + i\lambda \sigma _\nu
\mathord{\buildrel{\lower3pt\hbox{$\scriptscriptstyle\leftrightarrow$}}
\over \partial } _\mu  \bar \lambda  + i\theta \sigma _\mu
\mathord{\buildrel{\lower3pt\hbox{$\scriptscriptstyle\leftrightarrow$}}
\over \partial } _\nu  \bar \theta  + i\lambda \sigma _\mu
\mathord{\buildrel{\lower3pt\hbox{$\scriptscriptstyle\leftrightarrow$}}
\over \partial } _\nu  \bar \lambda  \\
  &+ (i\theta \sigma _b
\mathord{\buildrel{\lower3pt\hbox{$\scriptscriptstyle\leftrightarrow$}}
\over \partial } _\mu  \bar \theta  + i\lambda \sigma _b
\mathord{\buildrel{\lower3pt\hbox{$\scriptscriptstyle\leftrightarrow$}}
\over \partial } _\mu  \bar \lambda ) \cdot (i\theta \sigma ^b
\mathord{\buildrel{\lower3pt\hbox{$\scriptscriptstyle\leftrightarrow$}}
\over \partial } _\nu  \bar \theta  + i\lambda \sigma ^b
\mathord{\buildrel{\lower3pt\hbox{$\scriptscriptstyle\leftrightarrow$}}
\over \partial } _\nu  \bar \lambda ) \\
  &- (\partial _\mu  \phi  + i\partial _\mu  \theta \lambda  + i\theta
\partial _\mu  \lambda  - i\partial _\mu  \bar \theta \bar \lambda  -
i\bar \theta \partial _\mu  \bar \lambda ) \\
  &\cdot (\partial _\nu  \phi  + i\partial _\nu  \theta \lambda  + i\theta
\partial _\nu  \lambda  - i\partial _\nu  \bar \theta \bar \lambda  -
i\bar \theta \partial _\nu  \bar \lambda ))] ^2,
\end{align*}
in which the auxiliary intrinsic metric $\rho_{\mu \nu }$ can be eliminated by using its equation of motion. Here, $\rho ^{\mu \nu }$ is the inverse of the metric $\rho_{\mu \nu }$ with scale dimension $-2$, and $\rho$ stands for the determinant of $\rho_{\mu \nu }$, and $T$ stands for the brane tension. The part inside the square brackets has a scale dimension $-2$. It can be concluded that Eq.(27) is Weyl scale invariant under the transformation $\xi ^\mu \to e^d \xi^\mu$. Obviously, when the spinors are set to zero, it reduces to the Weyl scale invariant bosonic action [14]:
$$
I_B  =  - \frac{{f_s ^2 }}{2}\int {d^4 \xi } \sqrt {\left| {\det G}
\right|} [\frac{1}{4}G^{\mu \nu } \eta _{ab} \frac{{\partial x^a }}
{{\partial \xi ^\mu  }}\frac{{\partial x^b }}{{\partial \xi ^\nu  }}]^2 {\rm{ }}.
\eqno{(28)}
$$
The action (27) describes the effective oscillation modes of the brane into the bulk space, corresponding to the symmetry breakings in the $\theta ,\bar \theta ,\lambda ,\bar \lambda, \phi$ (super)space coordinates directions, and whose long wave length excitation modes are described by these Nambu-Goldstone fields associated with those broken symmetries, i.e., the symmetries related to $Q_\alpha  ,\bar Q_{\dot \alpha } ,S_\alpha  ,\bar S_{\dot \alpha } ,$ and $Z$.

   Given the higher nonpolynomial effective actions of Eq.(27) and (28), we notice that it is formidable to construct the associated canonical formalism as well as the study of their symmetries and quantum properties. Due to the high nonlinearity of such a theory, different approached has been investigated to find its canonical formalism [13, 14, 17]. Especially, in [13], an auxiliary scalar field with appropriate Weyl weight is introduced which transforms the nonlinear action into a quadratic action, for which the standard rules of the canonical analysis can be applied. The fully canonical analysis of the Weyl scale invariant $p=3$ brane action is beyond the scope of this paper and will be treated elsewhere.
\vspace{15pt}
\begin{flushleft}
{\Large 3. DISCUSSION AND SUMMARY}
\end{flushleft}
\vspace{15pt}

  In summary, in this paper, we start with the super Weyl
group and its automorphism subgroup. Then the embedded Weyl scale invariant $p=3$ brane breaks the super Weyl group symmetry in the target bulk space down to the $1+3$ dimensional Weyl $W(1,3)$ symmetry in the submanifold. The brane's dynamics is described by the Goldstone bosons (Goldstino fermions) modes associated with the broken spatial(Grassmann)
generators of the symmetry(supersymmetry) group.

\par On the other hand, if alternatively, the embedded $p=3$ brane breaks the target scale symmetry as well, therefore the new coset represensitive elements $$\Omega_s = e^{ix^\mu  p_\mu  } e^{i\phi (x)Z} e^{i[\theta (x)Q + \bar \theta (x)\bar Q + \lambda (x)S + \bar \lambda (x)\bar S]} e^{iu^\mu(x)K_\mu}e^{i\sigma(x)D}$$ lead to the following Cartan one-forms
$$
\Omega_s ^{ - 1} d\Omega_s = i(\omega_s ^a p_a  + \omega _{sQ}^\alpha  Q_\alpha   + \bar \omega _{ s \bar
Q\dot \alpha } \bar Q^{\dot \alpha }+\omega _{sZ} Z + \omega _{sS}^\alpha  S_\alpha   + \bar \omega _{ s \bar S\dot\alpha } \bar S^{\dot \alpha }  + \omega _{sk}^a K_a  + \omega _{sM}^{\mu \nu }M_{_{\mu \nu } }  + \omega _{sD} D),
\\
\eqno{(29)}
$$
in which $\omega_s ^a  = \omega^a e^{-\sigma}=dx^\mu e_{s\mu}^{\hspace{6pt} a}$ and $\omega _{sD}  = dx^\mu \partial_\mu \sigma=\omega_s ^a e_{sa}^{-1 \mu }\partial_\mu \sigma=\omega_s ^a D_\mu \sigma$, where the covariant derivative $D_\mu=e_{sa}^{-1 \mu }\partial_\mu$. Besides, considering Eq.(8), for a pure scale transformation, we find the dilaton field transforms as
$\sigma \to \sigma'=\sigma+d$. Accordingly, the dilaton field, which transforms with a shift, behaves as the Nambu--Goldstone field, signaling the sponetaneous breaking of the scale symmetry in the target bulk space. As a result, the effective brane action, which includes all the Nambu-Goldstone modes, is found to be
\begin{align*}
I &=  - T\int {d^4 x} \det e_s - \frac{T_s}{2}\int {d^4 x} \det e_s \eta^{ab} D_a \sigma D_b \sigma  \tag{30} \\
&=- T\int {d^4 x} e^{-4 \sigma} \det e -\frac{T_s}{2} \int {d^4 x} \det e e^{-2 \sigma} \eta^{ab} e_a^{-1 \mu}e_b^{-1 \nu} \partial_\mu \sigma \partial_\nu \sigma,
\end{align*}
in which $T_s$ is related to the broken scale of the dilatation symmetry. Expanding the first term explicitly, we then have the potential $V\propto e^{-4 \sigma}$. Hence, the VEV can be determined by estimating the value of $\left\langle e^{-4 \sigma}\right\rangle$, which becomes minimum when  $\left\langle \sigma \right\rangle$ goes to $\infty$. However, due to the unbound of the VEV of the dilaton field it follows that $\sigma$ cannot be a NG particle [26]. This fact thus indicates that it is infeasible to embed such a Minkosski brane, whose modes break simultaneously the dilatation symmetry and the supersymmetry in the target Super--Weyl bulk space. However, in the case of supergravity couplings this incompatibility can be eliminated.

  All in all, in this article, we present a $p=3$ brane scenario and show its dynamical origin of the Weyl scale invariant $p$--branes. It may be of special interest to understand the origin of the Weyl scale invariance for $p$--branes from a dynamical point of view. We also notice that in addition to the standard space time induced metric, the spinor fields modify the induced metric on the brane as well. In this sense these fields induce curvature on the brane. On the other hand, the coset approach has been extensively used to describe the spontaneous partial breaking of (extended) supersymmetry and construct actions of (super)brane dynamics[27-31]. It also has been applied to branes of $M$ theory with a large automorphism group of superalgebra [32]. In addition, in the theory of brane--world scenarios, the universe can be regarded as a four dimensional topological defect in the form of domain wall embedded in a higher dimensional spacetime [33]. Also, it has been shown that the non--BPS topological defects can be a source of possible SUSY breaking [34]. Actually, it is found that the $N=1$ supersymmetry preserved on the four dimensional world volume of one wall (brane) is completely broken by the coexistence of the other wall (brane)[35]. That scenario admits another possible origin of SUSY breaking, in contrast with the case of embedded branes, whose fluctuation modes break the bulk supersymmetry spontaneously.

\par The author thanks the THEP groups at Purdue and NITheP for supports.

\pagebreak

\begin{center}
{\bf REFERENCES}
\end{center}
\begin{description}

\item[1.] P. Brax and C. Bruck, Class.Quant.Grav.{\bf 20},R201(2003).

\item[2.] J. Polchinski, Phys. Rev. Lett. {\bf 75}, 4724 (1995).

\item[3.] A.Sen, JHEP {\bf 10}, 008(1999).

\item[4.] Matthias R. Gaberdiel,
Class.Quant.Grav.{\bf 17}, 3483(2000); A.Sen, hep-th/9904207

\item[5.] T. E. Clark, M. Nitta, and T. ter Veldhuis, Phys.Rev.D {\bf 70},
105005(2004)

\item[6.] M.B.Green and J.H.Schwarz, Phys.Lett. B {\bf 136}, 367(1984);
M.B.Green, J.H.Schwarz, and E.Witten, {\it Superstring Theory} (Cambridge Univ. Press, Cambridge, England, 1987) Vol.2.

\item[7.] M. Blagojevic, {\it Gravitation and Gauge Symmetries} (IOP Publishing, 2002).

\item[8.] Ryoyu Utiyama, Progress of Theor. Phys.{\bf 50}, 2080(1973); Progress of Theor.  Phys.{\bf 53}, 565(1975).

\item[9.] Kenji Hayashi and T.Kugo, Progress of Theor. Phys.{\bf 61}, 334(1979).

\item[10.] A. Bregman, Progress of Theor. Phys.{\bf 49}, 667(1973).

\item[11.] Kenji Hayashi, M.Kasuya, and T.Shirafuji, Progress of Theor. Phys.{\bf 57}, 431(1977).

\item[12.] A.Aurilia, A.Smailagic, and E.Spallucci, Phys.Rev.D {\bf 51},4410(1995).

\item[13.] J.Antonio Garcia, Roman Linares, and J.David Vergara, Phys. Lett. B {\bf 503}, 154(2001).

\item[14.] C.Alvear, R.Amorim, and J.Barcelos-Neto, Phys. Lett. B {\bf 273}, 415(1991).

\item[15.] M.S. Alves and J. Barcelos-Neto, Europhys.Lett.{\bf 7}, 395(1988), Erratum-ibid.{\bf 8},90(1989).

\item[16.] J.A. Nieto and Carmen A. Nunez, Nuovo Cim.B {\bf 106},1045(1991).

\item[17.] J.D. Vergara, 3rd Latin American Symposium On High-Energy Physics, PROCEEDINGS. Edited by E. Nardi. Bristol, IOP, 2000.

\item[18.] J.A.Nieto, Mod.Phys.Lett.A {\bf 16},2567(2001).

\item[19.] Lu-Xin Liu, Phys. Rev. D {\bf 74}, 45030(2006)[arXiv:hep-th/0602180].

\item[20.] H. Georgi, Phys.Rev.Lett.{\bf 98}, 221601(2007); Phys.Lett.B {\bf 650},275(2007)

\item[21.] Lu-Xin Liu, Mod. Phys. Lett. A {\bf 20},
2545(2005)[arXiv:hep-ph/0408210].

\item[22.] T.E. Clark, S. T. Love, and G. Wu, Phys.Rev.D {\bf 57},5912(1998).

\item[23.] T. E. Clark, M. Nitta, and T. ter Veldhuis, Phys.Rev.D {\bf 67},
085026(2003).

\item[24.] T. E. Clark, M. Nitta, and T. ter Veldhuis, Phys.Rev.D {\bf 70},
125011(2004).

\item[25.] E. A. Ivanov and V. I. Ogievetsky, Teor. Mat. Fiz. {\bf 25},
164(1975).

\item[26.] T.E. Clark and S.T. Love, Phys.Rev.D {\bf 61},057902(2000).

\item[27.] Lu-Xin Liu, Eur.Phys.J.C {\bf 62}, 615(2009),arXiv:0802.1299 [hep-th].

\item[28.] P. West, JHEP {\bf 0002}, 24(2000).

\item[29.] J. Hughes and J. Polchinski, Nucl. Phys. B {\bf 278}, 147(1986);
C. P. Burgess, E. Filotas, M. Klein, and F. Quevedo, JHEP {\bf 0310}, 41(2003); J. Bagger and A. Galperin, Phys. Lett. B {\bf 336}, 25(1994); Phys. Lett. B {\bf 412}, 296(1997); Phys. Rev. D {\bf 55}, 1091(1997).

\item[30.] S. Bellucci, E. Ivanov, and S. Krivonos, Phys. Lett. B {\bf 460},
348(1999); M. Rocek and A.A.Tseytlin, Phys. Rev. D {\bf 59}, 106001(1999); S. Bellucci, E. Ivanov, and S. Krivonos, Fortsch.Phys.{\bf 48}, 19(2000); F. Gonzalez-Rey, L. Y. Park, and M. Rocek, Nucl. Phys.B {\bf 544}, 243(1999).

\item[31.] I. N. McArthur, Hep-th/9908045; S. Bellucci, E. Ivanov, and S.
Krivonos, Phys.Lett.B {\bf 482}, 233(2000); S. Bellucci, E. Ivanov, and S. Krivonos, Nucl. Phys. Proc. Suppl.{\bf 102}, 26(2001); E. Ivanov, Theor. Math. Phys. {\bf 129}, 1543(2001).

\item[32.] O. Barwald and P. West, Phys. Lett. B {\bf 476}, 157(2000).

\item[33.] V.A.Rubakov and M.E.Shaposhnikov, Phys.Lett. B {\bf 125},
(1983)136; E.Papantonopoulos, Hep-th/0202044; H.A.Chamblin and H.S.Reall,
Nucl.Phys.B {\bf 562}, 133(1999).

\item[34.] G.Dvali and M.Shifman, Nucl.Phys.B {\bf 504}, 127(1997).

\item[35.] N.Maru, N.Sakai, Y.Sakamura, and R.Sugisaka,
Nucl.Phys.B {\bf 616}, 47(2001); Phys.Lett.B {\bf 496}, 98(2000); M.Eto, N.Maru, and N.Sakai, Nucl.Phys.B {\bf 696}, 3(2004).

\end{description}
\end{document}